\documentclass[final,3p,times,twocolumn]{elsarticle}

\usepackage{amssymb}

\usepackage{lineno}

\usepackage{hyperref}

\journal{Nuclear Instrumentations and Methods in Physics Research, Section A}

\begin{document}

\begin{frontmatter}



\title{Performance evaluation of the high-voltage CMOS active pixel sensor AstroPix for gamma-ray space telescopes}


\author[inst1]{Yusuke Suda}
\author[inst2]{Regina Caputo}
\author[inst2,inst3]{Amanda L. Steinhebel}
\author[inst4]{Nicolas Striebig}
\author[inst5]{Manoj Jadhav}
\author[inst1]{Yasushi Fukazawa}
\author[inst1]{Masaki Hashizume}
\author[inst2]{Carolyn Kierans}
\author[inst4]{Richard Leys}
\author[inst5]{Jessica Metcalfe}
\author[inst6]{Michela Negro}
\author[inst4]{Ivan Peri\'{c}}
\author[inst2]{Jeremy S. Perkins}
\author[inst7]{Taylor Shin}
\author[inst8,inst9]{Hiroyasu Tajima}
\author[inst2,inst3]{Daniel Violette}
\author[inst1]{Norito Nakano}

\affiliation[inst1]{organization={Physics Program, Graduate School of Advanced Science and Engineering, Hiroshima University},
            addressline={1-3-1 Kagamiyama}, 
            city={Higashihiroshima},
            postcode={739-8526}, 
            state={Hiroshima},
            country={Japan}}

\affiliation[inst2]{organization={NASA Goddard Space Flight Center},
            addressline={8800 Greenbelt Rd}, 
            city={Greenbelt},
            postcode={MD 20771}, 
            state={Maryland},
            country={USA}}

\affiliation[inst3]{organization={NASA Postdoctoral Program Fellow},
            addressline={8800 Greenbelt Rd}, 
            city={Greenbelt},
            postcode={MD 20771}, 
            state={Maryland},
            country={USA}}

\affiliation[inst4]{organization={ASIC and Detector Laboratory, Karlsruhe Institute of Technology},
            addressline={Hermann-von-Helmholtz-Platz 1}, 
            city={Karlsruhe},
            postcode={D-76344}, 
            state={Baden-Württemberg},
            country={Germany}}

\affiliation[inst5]{organization={Argonne National Laboratory},
            addressline={9700 S. Cass Avenue}, 
            city={Lemont},
            postcode={IL 60439}, 
            state={Illinois},
            country={USA}}

\affiliation[inst6]{organization={Department of Physics and Astronomy, Louisiana State University},
            addressline={202 Nicholson Hall}, 
            city={Baton Rouge},
            postcode={LA 70803}, 
            state={Louisiana},
            country={USA}}

\affiliation[inst7]{organization={University of California, Santa Cruz},
            addressline={1156 High Street}, 
            city={Santa Cruz},
            postcode={CA 95064}, 
            state={California},
            country={USA}}

\affiliation[inst8]{organization={Institute for Space–Earth Environmental Research, Nagoya University},
            addressline={Furo-cho, Chikusa-ku}, 
            city={Nagoya},
            postcode={464-8601}, 
            state={Aichi},
            country={Japan}}
            
\affiliation[inst9]{organization={Kobayashi-Maskawa Institute for the Origin of Particles and the Universe, Nagoya University},
            addressline={Furo-cho, Chikusa-ku}, 
            city={Nagoya},
            postcode={464-8602}, 
            state={Aichi},
            country={Japan}}

\begin{abstract}
AstroPix is a novel monolithic high-voltage CMOS active pixel sensor proposed for next generation medium-energy gamma-ray observatories like the All-sky Medium Energy Gamma-ray Observatory eXplorer (AMEGO-X).
For AMEGO-X AstroPix must maintain a power consumption of less than $1.5~\rm{mW/{cm}^2}$ while having a pixel pitch of up to $500~\rm{\mu m}$.
We developed the second and third versions of AstroPix, namely AstroPix2 and AstroPix3.
AstroPix2 and AstroPix3 exhibit power consumptions of $3.4~\rm{mW/{cm}^2}$ and $4.1~\rm{mW/{cm}^2}$, respectively.
While AstroPix2 has a pixel pitch of 250~$\rm{\mu m}$, AstroPix3 achieves the desired size for AMEGO-X with a pixel pitch of 500~$\rm{\mu m}$.
Performance evaluation of a single pixel in an AstroPix2 chip revealed a dynamic range from 13.9~keV to 59.5~keV, with the energy resolution meeting the AMEGO-X target value ($<10\%$ (FWHM) at 60 keV).
We performed energy calibration on most of the pixels in an AstroPix3 chip, yielding a mean energy resolution of 6.2~keV (FWHM) at 59.5~keV, with 44.4\% of the pixels satisfying the target value.
The dynamic range of AstroPix3 was assessed to span from 22.2~keV to 122.1~keV.
The expansion of the depletion layer aligns with expectations in both AstroPix2 and AstroPix3.
Furthermore, radiation tolerance testing was conducted on AstroPix.
An AstroPix2 chip was subjected to an equivalent exposure of approximately 10~Gy from a high-intensity $\rm{^{60}Co}$ source. 
The chip was fully operational after irradiation although a decrease in gain by approximately 4\% was observed.
\end{abstract}




\end{frontmatter}

\section{Introduction}
\label{sec:intro}
Cosmic gamma-ray observations in the MeV regime are essential to deepen our understanding of the physics in high energy astronomical phenomena such as gamma-ray bursts, active galactic nuclei, and others.
Since those transient phenomena happen at any direction in the Universe, all-sky MeV gamma-ray monitoring with a good localization capability will play an essential role in multi-messenger astronomy.

The All-sky Medium Energy Gamma-ray Observatory eXplorer (AMEGO-X) is one of the next generation MeV mission concepts~\cite{regina}.
Its gamma-ray telescope has an order of magnitude better sensitivity than previous missions in the energy range 100~keV to 1~GeV with the localization accuracy of $1^{\circ}$ ($90\%$ confidence level radius) for transient phenomena.
The gamma-ray telescope consists of four identical towers, each tower composed of a stacked silicon tracker and thallium-doped cesium iodide calorimeter module.
The tracker needs to accurately measure the positions of Compton scattering and the energy of recoil electrons for MeV photon detection.
To achieve the required sensitivity, the telescope needs a huge silicon area of about $24~\rm{m^2}$.
Due to limited satellite's power resource, there is a strict requirement in the power consumption of silicon sensors.
Thus, AMEGO-X demands a low power and low noise pixel silicon sensor with good position and energy resolution.

We have been developing a new type of monolithic active pixel silicon sensors, AstroPix, mainly for AMEGO-X.
In this work, we report the specifications and performance of the current iterations of AstroPix.
In Section~\ref{sec:astro}, we describe AstroPix and the specifications of the three different versions of AstroPix.
The results of basic performance evaluations of both the second version (AstroPix2) and the third version (AstroPix3) of AstroPix are described in Section~\ref{sec:v2} and \ref{sec:v3}, respectively.
We discuss the radiation tolerance of AstroPix2 in Section~\ref{sec:rad}.
The concluding remarks and future prospects are summarized in Section~\ref{sec:conc}.
\section{AstroPix}
\label{sec:astro}
AstroPix is a newly developed monolithic high voltage CMOS (HV-CMOS) active pixel sensor based on the experience of the developments of both ATLASPix and MuPix~\cite{peric,peric2}.
The technology node used is a 180~nm process.

\begin{figure*}[htb]
\begin{center}
\includegraphics[width=0.9\linewidth]{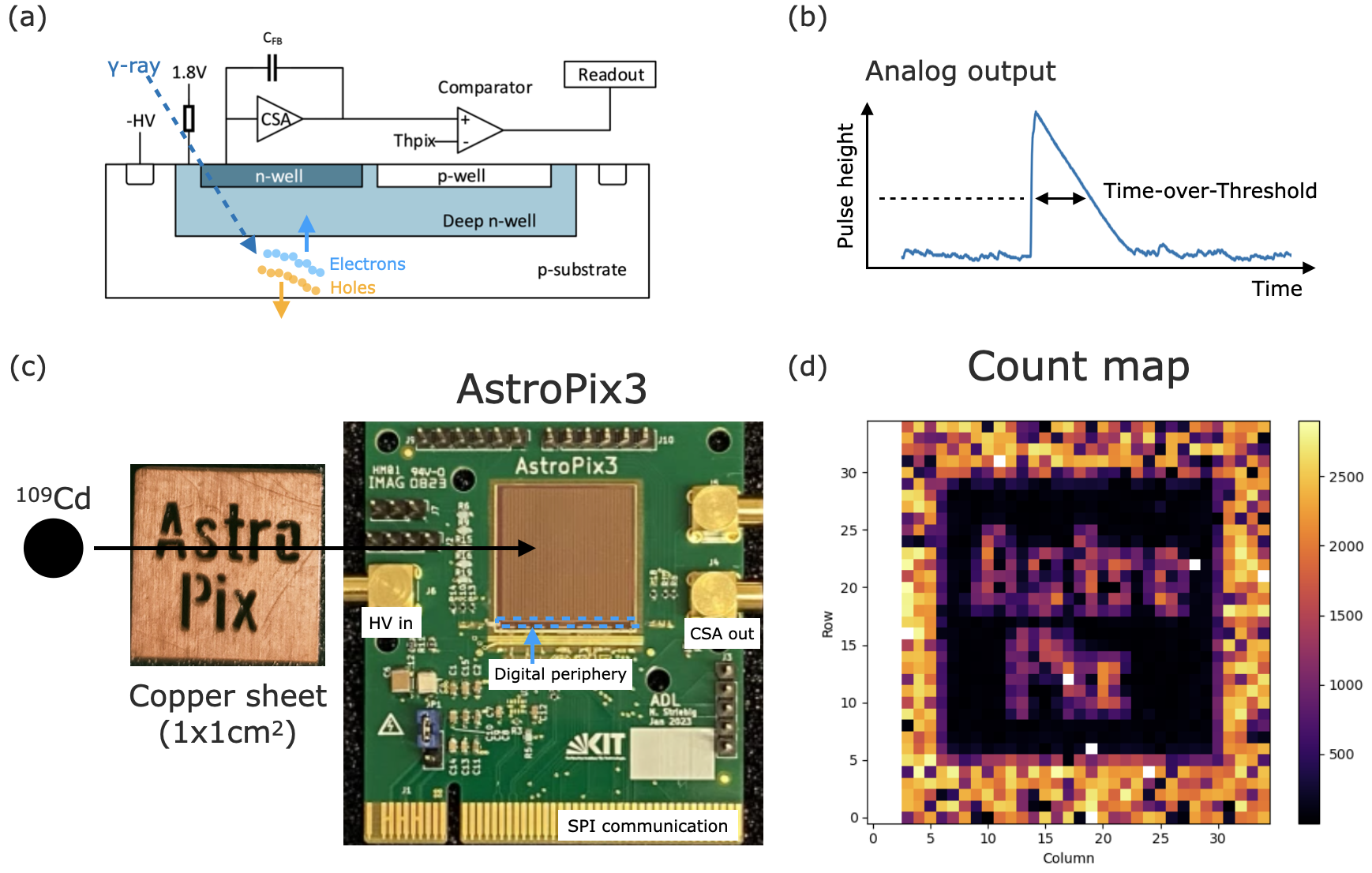}
\caption{
(a) Cross section of a single pixel in AstroPix sensor. The comparator output is processed in ``Readout'' section located at the digital periphery.
(b) The analog output of the charge-sensitive amplifier (CSA). 
(c) Photographs of the AstroPix3 chip on its carrier board and the copper mask with holes shaped like the letters ``AstroPix''.
(d) The count map of detected X-ray photons emitted from $\rm{^{109}Cd}$ through the copper mask placed between the chip and the radioisotope. The first three column pixels were masked due to their different amplifier configuration, which was intended for testing purposes. Eight noisy pixels in the remaining columns were masked, indicated by white dots.}
\label{fig:imaging}
\end{center}
\end{figure*}

The pixel CMOS circuits are implemented inside the deep N-well on the P-substrate (Fig.~\ref{fig:imaging} (a)).
By applying a high voltage (HV) between the N-well and P-substrate, the sensor layer can be fully depleted.
Electrons generated through gamma-ray interaction within the depletion volume are drifting towards and inducing a current signal on the N-well.
Subsequently, the signal undergoes processing on the pixel through a charge-sensitive amplifier (CSA) and comparator to trigger the digital hit processing logic.
Fig.~\ref{fig:imaging} (b) shows what the CSA output looks like. 
Finally, the time-over-threshold (ToT) values from all hit pixels are digitized on the periphery of the chip (a 12-bits ToT counter with a 200~MHz clock was used in this study).
Since AstroPix does not require any external analog readout electronics, AstroPix anticipates low noise hit probability and makes a low power silicon tracker feasible.
Moreover, significant reduction of passive materials inside the telescope, crucial for precise Compton imaging, can be achieved.

The goals for AstroPix to be adopted by AMEGO-X, based on the results from \cite{isabella}, are listed in Table~\ref{tab:req}.
\begin{table}[htbp]
  \caption{Goals for AstroPix. The energy resolution value is expressed in units of full width at half maximum (FWHM).}
  \label{tab:req}
  \centering
  \begin{tabular}{lc}
    \hline
    Power consumption & $<1.5~\rm{mW/{cm}^2}$\\
    Pixel pitch & $500\times500~\rm{\mu m^2}$\\
    Thickness & $500~\rm{\mu m}$\\
    Dynamic range & 25~keV - 700~keV\\
    Energy resolution & $<10\%$ (FWHM) at 60 keV\\
    \hline
  \end{tabular}
\end{table}
The first version of AstroPix (AstroPix1) chip ($5\times 5~\rm{mm^2}$) is $725~\rm{\mu m}$ thick and contains $18\times 18$ pixels whose pixel pitch is $175\times 175~\rm{\mu m}^2$.
It was used to understand the sensor and to develop the data acquisition tools.
The power consumption is $14.7~\rm{mW/{cm}^2}$.
The analog output from the CSA can only be read from the first row of pixels, which is also the case for both AstroPix2 and AstroPix3.
The emission lines from radioisotopes ranging from 13.9~keV to 122.1~keV can be observed in the analog output~\cite{spie}.
However, the digital data readout is not available due to a flaw in the chip design, which has since been understood and corrected.

The AstroPix2 chip ($10\times 10~\rm{mm^2}$) is $725~\rm{\mu m}$ thick and contains $35\times 35$ pixels whose pixel pitch is $250\times 250~\rm{\mu m^2}$.
The most significant upgrade from AstroPix1 is the ability to record digital data; it is now possible to read out the ToT value from each pixel across the chip.
The power consumption is $3.4~\rm{mW/{cm}^2}$.

The AstroPix3 chip ($19.5\times 18.6~\rm{mm^2}$), the first full reticle chip, is $725~\rm{\mu m}$ thick and contains $35\times 35$ pixels whose pixel pitch is $500\times 500~\rm{\mu m^2}$.
While expanding the pixel pitch to the target value, the implant size of the charge collecting electrode is kept as small as possible at $300\times 300~\rm{\mu m^2}$ to reduce capacitance.
The power consumption is $4.1~\rm{mW/{cm}^2}$ ($1.0~\rm{mW/{cm}^2}$ for the analog part).
Fig.~\ref{fig:imaging} (c) shows a photograph of the AstroPix3 chip on its carrier board and Fig.~\ref{fig:imaging} (d) represents the count map of X-rays emitted from $\rm{^{109}Cd}$ with a copper mask with holes shaped like the letters ``AstroPix''.
The shadow of the mask is visible, demonstrating a moderate imaging capability of AstroPix3.
\section{Performance evaluations of AstroPix2}
\label{sec:v2}
In this section, we present the results of basic performance evaluations of AstroPix2 chips. 
Although a higher resistivity wafer ($5\sim 20~\rm{k\Omega\cdot cm}$) is desirable to achieve the target depletion depth ($500~\rm{\mu m}$) and dynamic range, we tested AstroPix2 chips fabricated from a ($300\pm 100)~\rm{\Omega\cdot cm}$ resistivity wafer to perfect the design and testing tools in this work.
The nominal operating bias voltage was $-160~\rm{V}$ supplied by KEITHLEY 2450 SourceMeter.
The current draw was $\sim$5~nA at the nominal bias voltage, and the breakdown was observed when the bias voltage reaches at around $-190~\rm{V}$.

\begin{figure}[htb]
\begin{center}
\includegraphics[width=1\linewidth]{
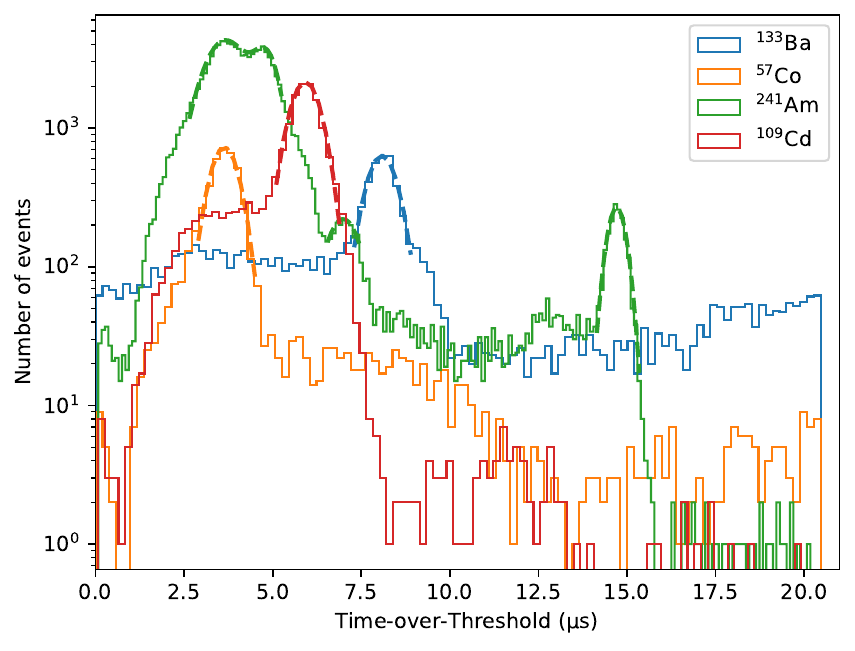}
\caption{Energy spectra of four radioisotopes ($\rm{^{133}Ba}$ in blue, $\rm{^{57}Co}$ in orange, $\rm{^{241}Am}$ in green, $\rm{^{109}Cd}$ in red) in units of ToT obtained from a single pixel of AstroPix2. The curves represent Gaussian fitted curves for each photopeak ($\rm{^{133}Ba}$: 31.0~keV. $\rm{^{57}Co}$: 14.4~keV. $\rm{^{241}Am}$ 13.9~keV, 17.8~keV, 26.3~keV, and 59.5~keV. $\rm{^{109}Cd}$: 22.2~keV).}
\label{fig:v2spec}
\end{center}
\end{figure}

\begin{figure}[htb]
\begin{center}
\includegraphics[width=0.9\linewidth]{
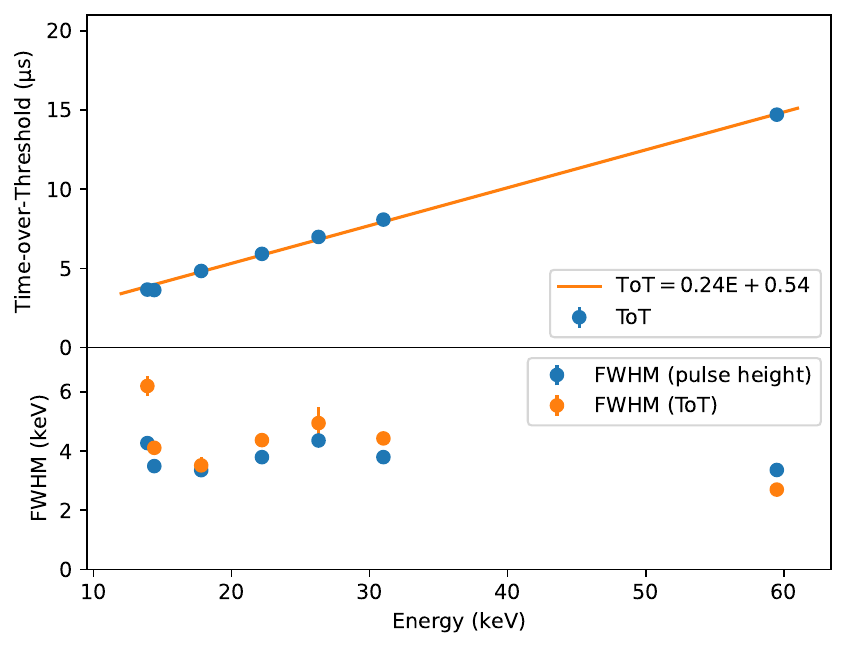}
\caption{Photopeak ToT value and energy resolution as a function of energy for a single pixel of AstroPix2. The energy calibration was conducted in the range from 13.9~keV to 59.5~keV using a linear function (solid line), and the resultant values are indicated in the legend.}
\label{fig:v2calib}
\end{center}
\end{figure}

\begin{figure}[htb]
\begin{center}
\includegraphics[width=0.9\linewidth]{
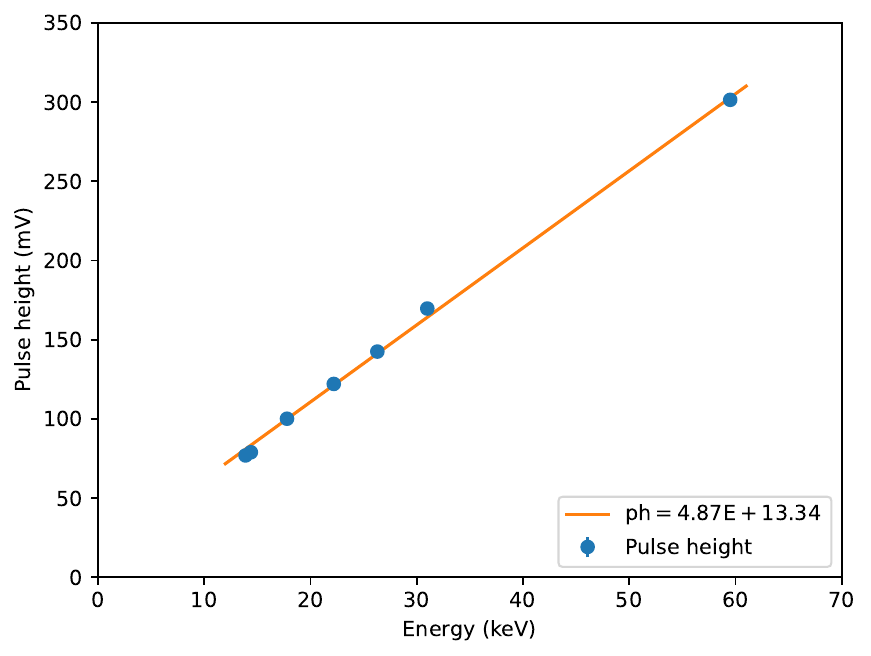}
\caption{Pulse height of the CSA output as a function of energy for a single pixel of AstroPix2.}
\label{fig:v2ph}
\end{center}
\end{figure}

Fig.~\ref{fig:v2spec} shows the energy spectra of four radioisotopes ($\rm{^{133}Ba}$, $\rm{^{57}Co}$, $\rm{^{241}Am}$, and $\rm{^{109}Cd}$) in units of ToT obtained from a single pixel.
The emission lines ranging from 13.9~keV to 59.5~keV are clearly visible.
Using the Gaussian fitted peak ToT values, the energy calibration of the same pixel was performed as shown in Fig.~\ref{fig:v2calib}.
The data points can be fitted by a linear function.
The dynamic range in the current configuration is from 13.9~keV to 59.5~keV (or 80~keV, assuming the fitted function can be applied until the end of the ToT counter bit).
Fig.~\ref{fig:v2ph} shows the correlation between the CSA output and energy, obtained from a single pixel in a different AstroPix2 chip.

The energy resolutions of each photopeak are also presented in the bottom panel of Fig.~\ref{fig:v2calib}.
The pulse heights of the CSA output from this pixel were measured by a multi-channel analyzer, simultaneously with the ToT measurements, to estimate spectral performance in the analog output.
The energy resolutions in digital (ToT) are comparable to those in analog and generally meet the target value.

\begin{figure}[htb]
\begin{center}
\includegraphics[width=0.9\linewidth]{
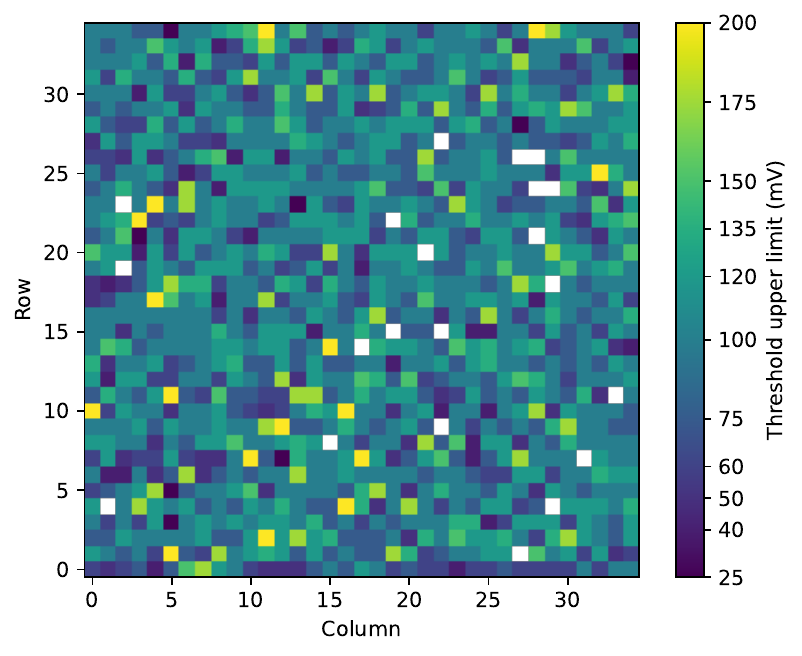}
\caption{Map of the threshold upper limit obtained from an AstroPix2 chip. White dots represent noisy pixels that were not used in this study.}
\label{fig:v2ulmap}
\end{center}
\end{figure}

\begin{figure}[htb]
\begin{center}
\includegraphics[width=0.9\linewidth]{
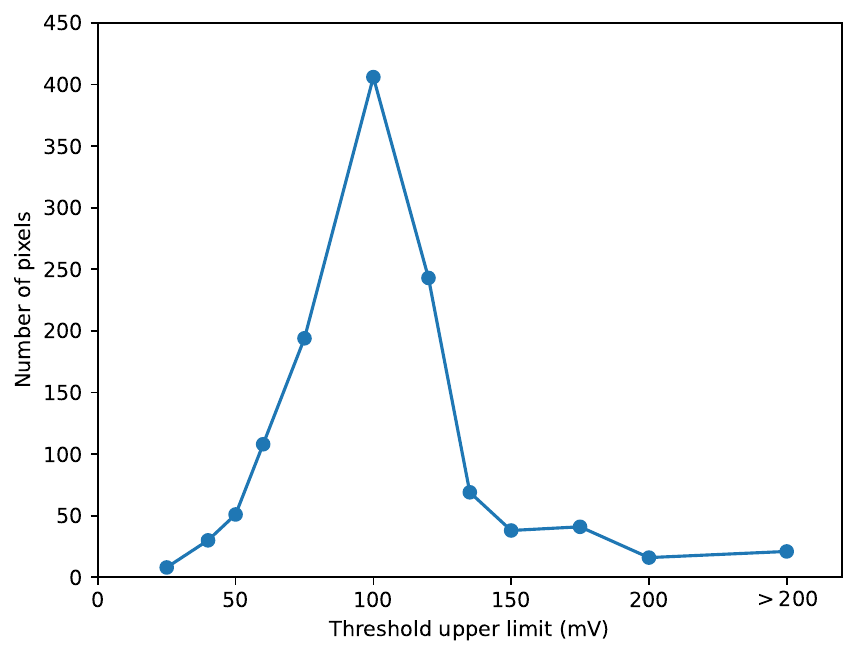}
\caption{Threshold upper limit distribution obtained from an AstroPix2 chip. Approximately 90\% of the pixels have their thresholds below 135~mV which corresponds to the estimated energy of 25~keV for the calibrated pixel shown in Fig.~\ref{fig:v2ph}.}
\label{fig:v2uldist}
\end{center}
\end{figure}

To evaluate the overall noise level across the chip, hit rates were measured with various threshold voltages in the same chip used in Fig.~\ref{fig:v2ph}.
Only one pixel was enabled to read out the digital data at a time in order to minimize dead time.
We define here the noise level (threshold upper limit) as the threshold at which the noise hit rate (number of hits in 30~sec) becomes zero for the first time when starting from the lowest threshold (25~mV).
The threshold scan was performed with the numbers shown in the color bar of Fig.~\ref{fig:v2ulmap}. 
The estimated noise level varies among pixels with no clear pattern nor structure observed across the chip, as shown in Fig.~\ref{fig:v2ulmap}.
Assuming the gain variation is not large, one can utilize the linear function shown in Fig.~\ref{fig:v2ph} to estimate the fraction of pixels that satisfy the lower edge of the target dynamic range.
It is estimated that nearly 90\% of the pixels meet the target value of 25~keV in the current configuration (Fig.~\ref{fig:v2uldist}).

\begin{figure}[htb]
\begin{center}
\includegraphics[width=1\linewidth]{
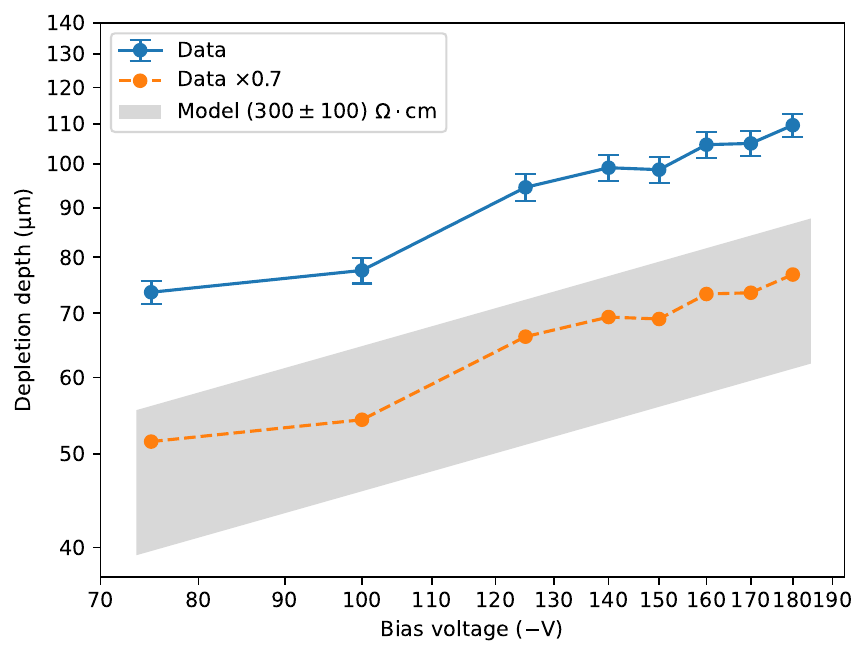}
\caption{Bias voltage dependence of the depletion depth measured in a single pixel of AstroPix2. The blue points are the measured depletion depths. The error bars represent $1~\sigma$ statistical errors. The orange points are the scaled data to be within the PN junction model band in gray. }
\label{fig:v2dep}
\end{center}
\end{figure}

\begin{figure}[htb]
\begin{center}
\includegraphics[width=0.9\linewidth]{
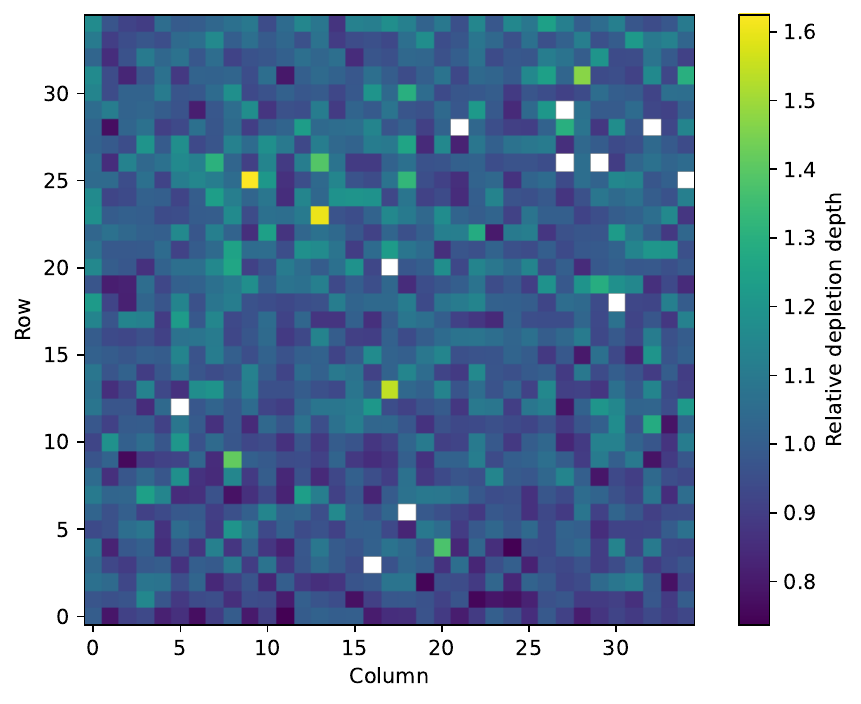}
\caption{Map of the relative depletion depth obtained from an AstroPix2 chip. The used bias voltage is $-160~\rm{V}$. White dots represent noisy pixels that were not used in this study.}
\label{fig:v2depmap}
\end{center}
\end{figure}

\begin{figure}[htb]
\begin{center}
\includegraphics[width=0.9\linewidth]{
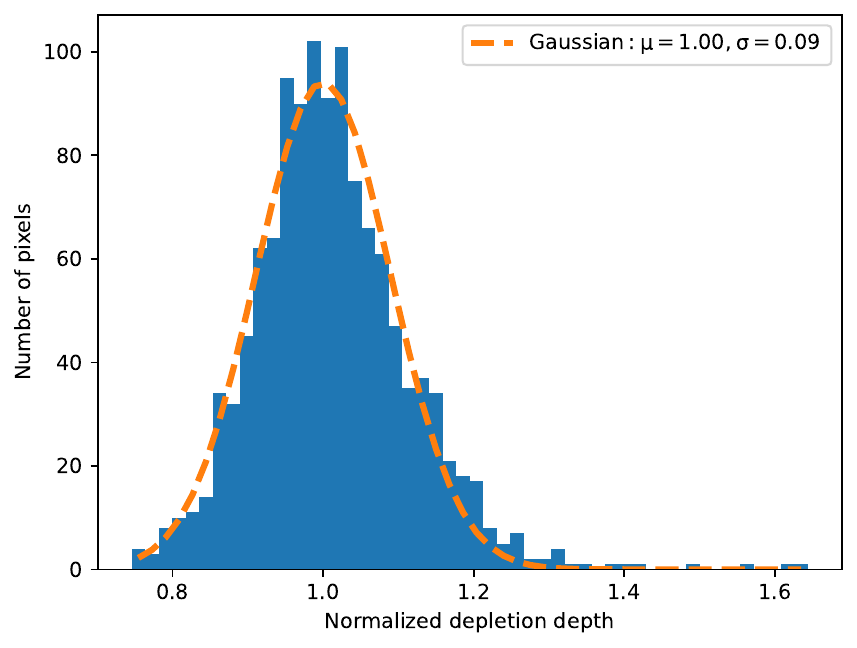}
\caption{Normalized depletion depth distribution obtained from an AstroPix2 chip. The dotted curve represents the fitted Gaussian curve with the sigma of 0.09.}
\label{fig:v2depdist}
\end{center}
\end{figure}

The bias voltage dependence of the depletion depth in a single pixel was derived as follows:
\begin{enumerate}
    \item Measure the event rate of 59.5~keV photons emitted by $\rm{^{241}Am}$ located at a distance $L$ of 3~cm from the chip.
    \item Scan the bias voltages and do the step 1 at each stage.
    \item Calculate the depletion depth by using the event rate, photon emission rate, solid angle, and photoelectric cross section.
    We assume the radioisotope to be a point-like source, meaning the solid angle is defined by the pixel pitch area ($250\times 250~\rm{\mu m^2}$) divided by $4\pi L^2$.
    \item Compare the data with a simple PN junction model given by the following equation.
        \begin{displaymath}
            d = \sqrt{2\epsilon\mu\rho\left(V_{\rm{bias}} + V_{\rm{built\_in}} \right)}
        \label{eq:dep}
        \end{displaymath}
        where $d$ is the theoretical depletion depth, $\epsilon$ the permittivity, $\mu$ the hole mobility, $\rho$ resistivity, $V_{\rm{bias}}$ the bias voltage, and $V_{\rm{built\_in}}$ the built-in potential.
\end{enumerate}
Fig.~\ref{fig:v2dep} shows the bias voltage dependence of the depletion depth in the same pixel as in Fig.~\ref{fig:v2spec} and \ref{fig:v2calib}, located at the middle of the first row pixels.
Data points can be within the uncertainty range of the model if a scale factor of 0.7 applied.
This factor reflects the systematic uncertainty due to the use of this method for photon counting because it could not subtract lower energy photons from the photon counting region.
Nevertheless, the obtained curve has a similar tendency as the model, implying the depletion layer of the AstroPix2 chip grows as expected.
In order to fully deplete, it is evident that we need to utilize a chip substrate with a much higher resistivity as expected.
The full scan of the depletion depth over the chip at the nominal bias voltage was performed, and no clear pattern nor structure is visible as shown in Fig.~\ref{fig:v2depmap}.
Fig.~\ref{fig:v2depdist} shows the normalized depletion depth distribution.
Using the fitted sigma from that distribution, the deviation of the measured depletion depths is estimated to be 9\%.
\section{Performance evaluations of AstroPix3}
\label{sec:v3}
We used an AstroPix3 chip fabricated from a ($300\pm 100)~\rm{\Omega\cdot cm}$ resistivity wafer for the evaluation of basic performance.
The first three column pixels were not used since their amplifier configuration is different for testing purposes.
The current draw was $\sim$50~nA at the nominal bias voltage of $-350~\rm{V}$, and the breakdown was observed when the bias voltage reaches at around $-400~\rm{V}$.
This higher breakdown voltage is attributed to two factors: the wider clearance between the outer pixel deep N-well and the chip edge in AstroPix3 ($600~\rm{\mu m}$ at minimum) compared to that in AstroPix2 ($150~\rm{\mu m}$), and the wider spacing between the inter-pixel guard ring and the inner deep N-well metal guard ring in AstroPix3 ($\sim$100~$\rm{\mu m}$) compared to that in AstroPix2 ($2~\rm{\mu m}$).

\begin{figure}[htb]
\begin{center}
\includegraphics[width=1\linewidth]{
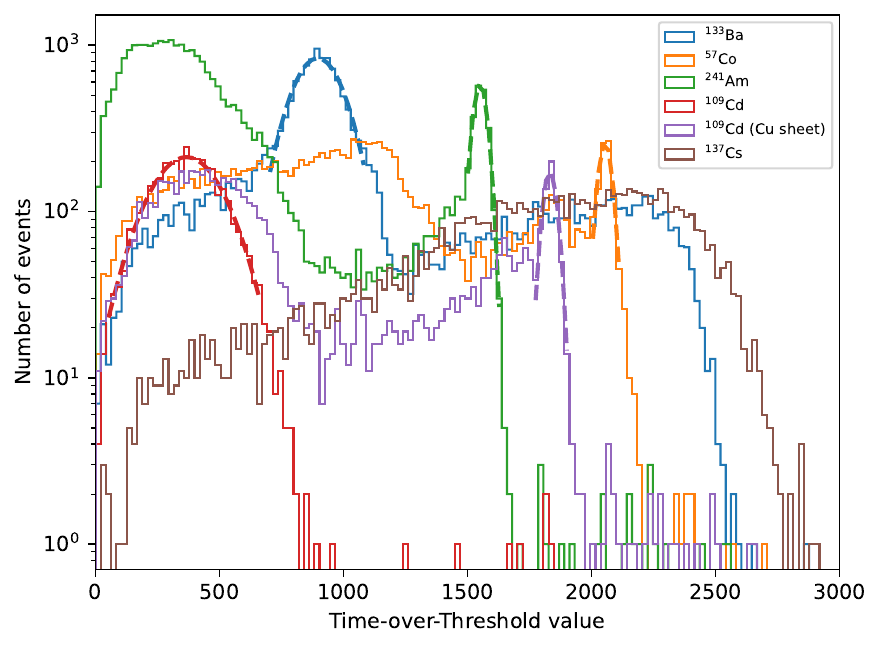}
\caption{Energy spectra in units of ToT obtained from a single pixel of AstroPix3 ($\rm{^{133}Ba}$ in blue, $\rm{^{57}Co}$ in orange, $\rm{^{241}Am}$ in green, $\rm{^{109}Cd}$ in red, and $\rm{^{137}Cs}$ in brown). For the purple spectrum, the data was taken with a copper sheet placed between $\rm{^{109}Cd}$ and the sensor to reduce the hit rate due to lower energy photons.}
\label{fig:v3spec}
\end{center}
\end{figure}

The energy spectra of five radioisotopes ($\rm{^{133}Ba}$, $\rm{^{57}Co}$, $\rm{^{241}Am}$, $\rm{^{109}Cd}$, and $\rm{^{137}Cs}$) in units of ToT obtained from a single pixel is shown in Fig.~\ref{fig:v3spec}.
Five photopeaks (22.2~keV, 31.0~keV, 59.5~keV, 88.0~keV, and 122.1~keV) are visible, but three photopeaks emitted from $\rm{^{241}Am}$ (13.9~keV, 17.8~keV, and 26.3~keV) cannot be resolved.
The increase in the implant size of the charge collecting electrode can be one of the reasons for the deterioration of energy resolution compared to AstroPix2.
The saturation of the CSA output can be observed in the $\rm{^{137}Cs}$ spectrum.
The higher ToT component in $\rm{^{133}Ba}$ could be Compton edge for higher energy photons, such as 302.9~keV and 356.0~keV (the Compton edges for those photons are 164.3~keV and 207.3~keV, respectively).

\begin{figure}[htb]
\begin{center}
\includegraphics[width=1\linewidth]{
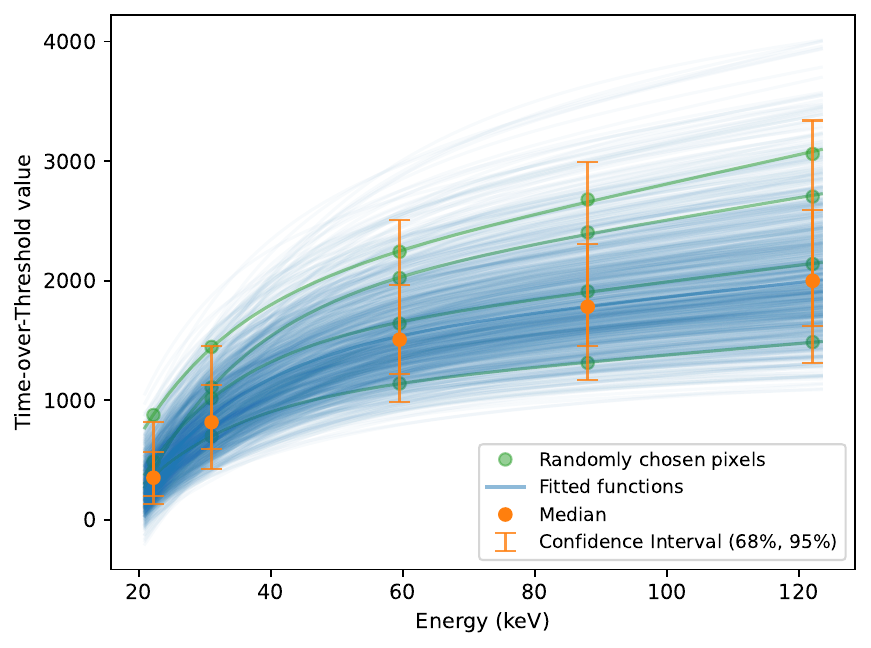}
\caption{Full energy calibration curves for an AstroPix3 chip. Green points and curves are the Gaussian fitted peak ToT values of randomly chosen pixels and their fitted calibration curves, respectively. Blue curves show calibration curves for all other pixels. Orange points and error bars represent medians of peak ToT distribution and their confidence intervals (68\% and 95\%).}
\label{fig:v3calib}
\end{center}
\end{figure}

\begin{figure}[htb]
\begin{center}
\includegraphics[width=1\linewidth]{
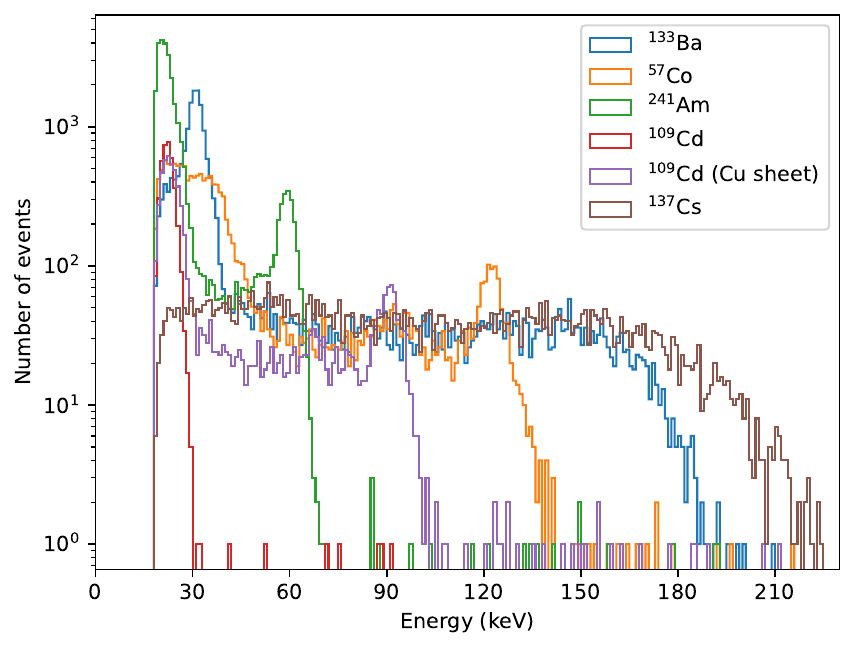}
\caption{Energy spectra obtained from a single pixel of AstroPix3.}
\label{fig:v3enespec}
\end{center}
\end{figure}

Full energy calibration over the sensor was performed (eight noisy pixels were masked as shown in Fig.~\ref{fig:imaging}(d). 1112 pixels were tested in total).
The fraction of pixels with 22.2~keV photopeak is 92.4\%, which means that at least 92.4\% of the tested pixels meet the goal value of the lower limit of the dynamic range (25~keV).
We selected pixels having the five photopeaks, and 89.6\% of the tested pixels were calibrated.
Fig.~\ref{fig:v3calib} shows the obtained calibration curves.
Here, we use an empirical function ($ToT = aE+b[1-\exp(-E/c)]+d$, where $E$ is the photopeak energy) to fit the data points (the median and 68\% confidence interval of the fitted parameters are $a=5^{+3}_{-2}$, $b=3775^{+1483}_{-916}$, $c=19^{+6}_{-5}$, and $d=-2328^{+755}_{-1248}$).
The spectra in Fig.~\ref{fig:v3spec} were calibrated using the fitted function, as presented in Fig.~\ref{fig:v3enespec}.
The five photopeaks were successfully reconstructed.
Therefore, the dynamic range of AstroPix3 in the current configuration is evaluated as from 22~keV to 122.1~keV (or approximately 200~keV, assuming the calibration curves are also applicable at energies above 122.1~keV).

\begin{figure}[htb]
\begin{center}
\includegraphics[width=1\linewidth]{
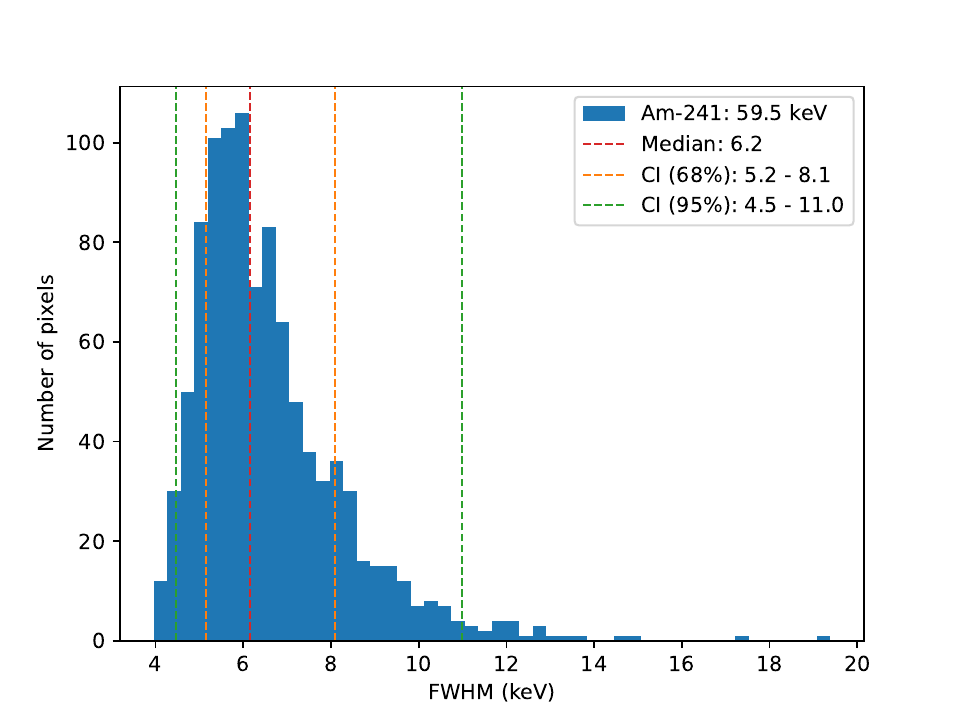}
\caption{Energy resolution distribution of 59.5~keV photons ($\rm{^{241}Am}$) obtained from an AstroPix3 chip. CI stands for confidence interval.}
\label{fig:v3eneres}
\end{center}
\end{figure}

The median of the energy resolution distribution for 59.5~keV photopeak is 6.2~keV (FWHM), and 44.4\% of the calibrated pixels meet the target value, as shown in Fig.~\ref{fig:v3eneres}.

We employ a similar methodology as described in Section~\ref{sec:v2} to evaluate the depletion depth of AstroPix3, with the following modifications.
First, a copper sheet was inserted between the chip and $\rm{^{241}Am}$ to filter out lower energy photons while allowing 59.5~keV photons.
By using this filter, we can reduce the data acquisition rate and investigate the detector response to monochromatic hard X-ray photons, making it easier to extract the photopeak events from the spectrum.
Subsequently, the resulting spectra were fitted using a combined function of three components: 1. primary Gaussian for the photopeak of 59.5~keV photons, 2. secondary Gaussian for the Compton back-scattered peak around 50~keV, and 3. error function for the lower energy flat spectral component.
Then, the photopeak (Gaussian-like) distributions were obtained by subtracting the two extra components from the spectra.
Finally, the number of the photopeak events was calculated by integrating the subtracted spectrum.
According to Geant4-based simulations, a non-negligible amount of the photoelectric events does not deposit all their energies in the depletion layer due to the expected depletion thickness at the given bias voltage not being thick enough.
To address this, we correct the data points by introducing an ``effective volume,'' in which photoelectric electrons deposit all their energies in the depletion layer (the correction factor to the nominal volume varies from 0.80 to 0.89 depending on the bias voltage), and an ``effective time'' to account for the dead time in the readout (the correction factor to the observed time is around 0.95).

\begin{figure}[htb]
\begin{center}
\includegraphics[width=1\linewidth]{
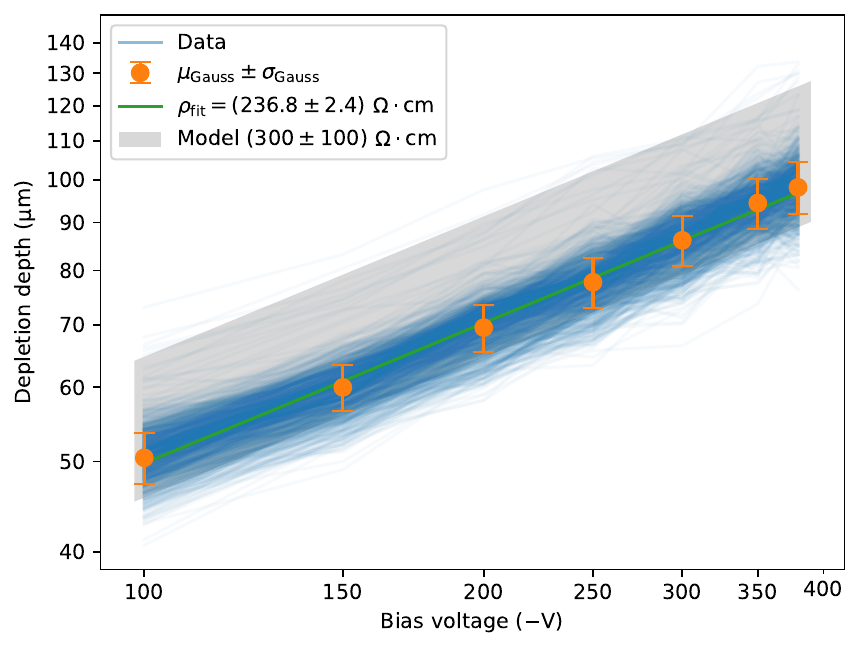}
\caption{Bias voltage dependence of the depletion depths for an AstroPix3 chip. The blue lines indicate the measured depletion depths for each pixel. The orange points show the Gaussian means of depletion depth distribution and the error bars are the fitted sigmas ($50\pm3, 60\pm3, 69\pm4, 78\pm5, 86\pm5, 94\pm6, 98\pm6~\rm{\mu m}$ at 100, 150, 200, 250, 300, 350, 380 V, respectively). The green line is the fitted PN junction model, resulting the fitted resistivity of $(236.8\pm 2.4)~\rm{\Omega\cdot cm}$. The gray band represents the PN junction model band with the uncertainty in the resistivity.}
\label{fig:v3dep}
\end{center}
\end{figure}

Fig.~\ref{fig:v3dep} shows the measured depletion depths as a function of bias voltage, and the model curve with the uncertainty in the resistivity.
The moderate agreement between the data and model can be seen, including the shape of the curves.
We note that there are several systematic uncertainties in this study, such as the uncertainty in the activity of $\rm{^{241}Am}$, and the uncertainty due to the assumption about the detection volume ($500~\rm{\mu m}\times 500~\rm{\mu m}\times\rm{depletion~depth}$).
Similar to AstroPix2, the depletion depth of AstroPix3 develops as expected, but it is again evident that we need to develop the chip with a higher resistivity wafer.
\section{Radiation tolerance}
\label{sec:rad}
\begin{figure}[htb]
\begin{center}
\includegraphics[width=0.9\linewidth]{
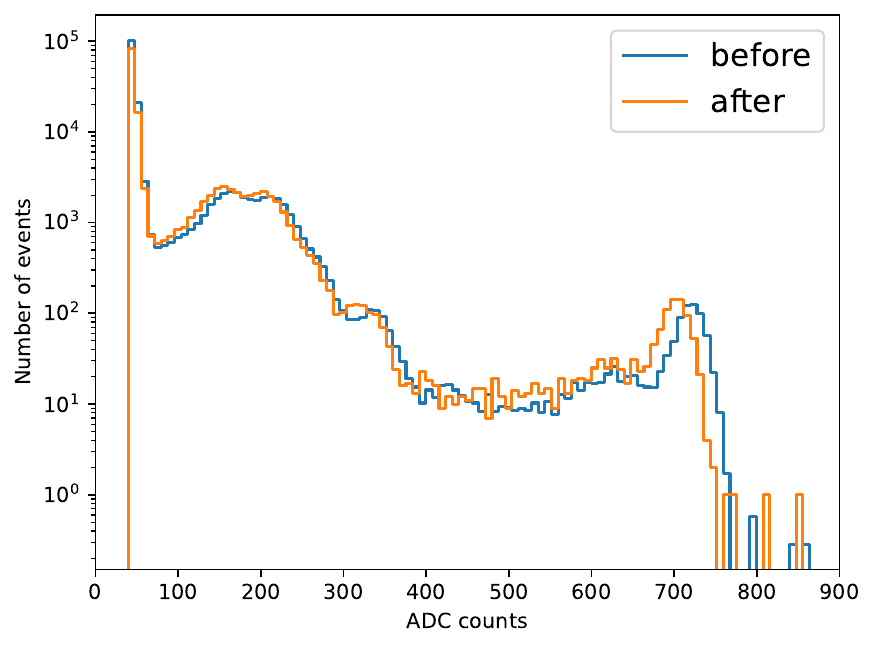}
\caption{Energy spectra of $\rm{^{241}Am}$ before (blue) and after (orange) irradiation obtained from a single pixel of AstroPix2. The measurements were performed with a multi-channel analyzer by recording pulse height of the CSA output, and the horizontal axis represents the analog-to-digital converter (ADC) value. The blue spectrum was normalized by the measurement time with respect to the orange one.}
\label{fig:v2rad}
\end{center}
\end{figure}

Since we intend to deploy AstroPix in space, it is crucial to assess the radiation tolerance of the sensor.
To this end, we exposed an AstroPix2 chip to a high-intensity $\rm{^{60}Co}$ source at Hiroshima University~\cite{co60}.
The test was performed at room temperature and the chip was powered during irradiation.
The estimated irradiation dose for this test is $\sim$10~Gy (the dose rate was 0.024~Gy/s), which is roughly equivalent to the expected value per year in orbit.
Following the irradiation, the chip functioned normally, albeit with a 60\% increase in the HV bias current draw.
By comparing the energy spectra taken before and after irradiation, a slight decrease in gain ($\sim$4\%) was observed, but without any significant deterioration in noise.
An example of the comparison of energy spectra is shown in Fig.~\ref{fig:v2rad}.
Further irradiation tests on this chip, as well as an AstroPix3 chip, are planned to more accurately evaluate the radiation tolerance of AstroPix.
\section{Conclusions and Outlook}
\label{sec:conc}
For future MeV gamma-ray space telescopes, we developed both AstroPix2 and AstroPix3, and conducted performance evaluations of the fabricated chips.
AstroPix2 is the first version in the AstroPix series equipped with proper digital readout functionality, demonstrating similar spectral performance to that of the analog output from the CSA.
The dynamic range in a single pixel of AstroPix2 was estimated to be from 13.9~keV to 59.5~keV, and the energy resolution essentially meets the target value for AMEGO-X.
AstroPix3 was developed to fulfill the target value of the pixel pitch.
The power consumption of $4.1~\rm{mW/{cm}^2}$ exceeds the target value by a factor of 3.
An improved time stamp generation and readout architecture are implemented in the fourth version of AstroPix (AstroPix4), which will reduce power consumption~\cite{nicolas}.
Large deviations in peak ToT values found in AstroPix3 will be suppressed in AstroPix4, as it is equipped with a functionality for pixel-by-pixel threshold tuning.
The dynamic range of AstroPix3 was evaluated to be from 22.2~keV to 122.1~keV (or around 200~keV, assuming the calibration curves are also applicable at energies higher than 122.1~keV).
In future versions of AstroPix, the upper limit of the dynamic range will reach to its target value by introducing a dynamic feedback capacitance in the CSA.
Despite the increase in pixel pitch, the mean energy resolution at 59.5~keV was measured to be 6.2~keV (FWHM), with 44.4\% of the pixels meeting the target value.
In both the cases of AstroPix2 and AstroPix3, the depletion layer appears to expand as expected.
To achieve the target depletion depth of 500~$\rm{\mu m}$, the development and evaluation of AstroPix chip fabricated from a higher resistivity wafer are necessary.
After irradiation with $\sim$10~Gy of $\rm{^{60}Co}$, the AstroPix2 chip was fully operational although a decrease in gain of $\sim$4\% was observed before and after irradtiaion.
\section{Acknowledgments}
\label{sec:ack}
The authors gratefully acknowledge financial support from NASA (18-APRA18-0084).
This work was supported by JSPS KAKENHI Grant Numbers JP23K13127 and JP23H04897.
We would like to thank staff in the radiation facility of the School of Engineering, Hiroshima University, for granting us access to their high-intensity $\rm{^{60}Co}$ facility.
The authors would like to thank the anonymous journal referees for their feedback, which helped to improve the quality of this paper.









\end{document}